  %%%%%%%This requires the PHYZZX.TEX macropackage

 \tolerance=10000
\input phyzzx

 \def\unit{\hbox to 3.3pt{\hskip1.3pt \vrule height 7pt width .4pt \hskip.7pt
\vrule height 7.85pt width .4pt \kern-2.4pt
\hrulefill \kern-3pt
\raise 4pt\hbox{\char'40}}}

\def\nup#1({Nucl.\ Phys.\  {\bf B#1}\ (}

\def\bb{{\bar \beta}}
%%%%%%%%%%%%%%%%%%%%%%%%%%%%%%%%%%%%%%%%%%%%%%%%%%%%%%%%%%%%%%%%%%%%
\REF\Oo{H. Ooguri and C. Vafa, Mod. Phys. Lett. {\bf A5} (1990) 1389; Nucl.
Phys. {\bf 361} (191)
469.}
\REF\GHR{S.J. Gates Jr., C.M. Hull and M. Ro\v cek, Nucl. Phys. {\bf B248}
(1984)  157.}
\REF\Kou{E. Kiritsis, C. Kounnas and D. Lust, Int. J. Mod. Phys. {\bf A9}
(1994) 1361.}
\REF\Bus{T.H. Buscher, Phys. Lett. {\bf 159B} (1985) 127.}
\REF\Hsig{ C.M. Hull, Nucl. Phys. {\bf B267} (1986)  266.}
\REF\Hsiga{ C.M. Hull,  Phys. Lett. {\bf 178B} (1986) 357.}
\REF\Hsigb{ C.M. Hull, in the Proceedings of the First Torino Meeting on
Superunification and Extra Dimensions,  edited by R. D'Auria and P. Fr\' e,
(World Scientific, Singapore, 1986).}
\REF\Van{ C.M. Hull, in Super Field Theories  (Plenum, New York, 1988), edited
by H. Lee and G. Kunstatter.}
\REF\Roc{M. Ro\v cek, in {\it Essays on Mirror Manifolds}, edited by S.-T. Yau,
International Press, Hong Kong, 1992.}
\REF\HW{C.M. Hull and E. Witten, Phys.Lett. {\bf 160B} (1985)  398.}
\REF\HT{C.M. Hull and P.K. Townsend, Nucl. Phys. {\bf B274} (1986)  349.}
\REF\GHL{J. Grundberg, A. Karlhede, U. Lindstrom, G. Theodoridis,  Nucl.Phys.
{\bf B282} (1987) 142 ; Class.Quant.Grav. {\bf 3L}  (1986) 129;  I. Jack,
Nucl. Phys.  {\bf B371}  (1992) 482.}
\REF\RSS{M. Ro\v cek, K. Schoutens and A. Sevrin, Phys. Lett. {\bf 265B} (1991)
303.}
\REF\HP{P.S. Howe and G. Papadopoulos, Nucl. Phys. {\bf B289} (1987) 264.}
\REF\NS{D. Nemeschansky and A. Sen, Phys. Lett. {\bf  178B} (1986) 385.}
\REF\Gates{  S.J.Gates, Jr. and H.Nishino, Mod.Phys.Lett. {\bf A7}, (1992)
      2543; S.J. Gates, Jr., S.V.Ketov and H. Nishino, Phys.Lett. {\bf 307B}
     (1993) 323;  Phys.Lett. {\bf 297B}
     (1993) 99; Nucl.Phys. {\bf B393}
     (1993) 149 and Phys.Lett. {\bf 307B}
     (1993) 331.}
\REF\Sez{E. Bergshoeff and E. Sezgin, Phys. Lett.
{\bf B292} (1992) 87; E. Sezgin, hep-th/9212092,  in proceedings of  Trieste
Summer Workshop on Superstrings and Related Topics, 1992.}
\REF\MK{D. Kutasov and E. Martinec, hep-th/9602049.}
%%%%%%%%%%%%%%%%%%%%%%%%%%%%%%%%%%%%%%%%%%%%%%%%%%%%%%%%%%%%%%%%%%%%

\Pubnum{ \vbox{ \hbox {QMW-96-13}  \hbox{hep-th/9606190}} }
\pubtype{}
\date{June, 1996}

\titlepage

\title {\bf  The Geometry of $N=2$ Strings with Torsion}

\author{C.M. Hull}
\address{Physics Department,
Queen Mary and Westfield College,
\break
Mile End Road, London E1 4NS, U.K.}
\vskip 0.5cm

\abstract {$N=2$ string theories are formulated in space-times with 2 space and
2 time
dimensions. If the world-sheet matter system consists of 2 chiral superfields,
the space-time is
Kahler and the dynamics are those of anti-self-dual gravity. If instead one
chiral superfield and one
twisted chiral superfield are used, the space-time is a hermitian manifold with
torsion and a
dilaton. The string spectrum consists of a scalar, which is a potential $K$
determining the metric,
torsion and dilaton.  The dynamics imply that the curvature with torsion is
anti-self-dual, and an
action is found for the potential $K$. It is argued that any $N=2$  sigma-model
with twisted
chiral multiplets in any dimension can be deformed to a conformally invariant
theory if the lowest
order contribution to the conformal anomaly vanishes.  If there are isometries,
more general geometries are possible in which the   dilaton is  the Killing
potential for a holomorphic Killing vector.}

\endpage

In [\Oo] a superstring with N=2  supersymmetry was shown to describe self-dual
gravity
 in a Kahler space-time with signature (2,2) (or (4,0)). The world-sheet matter
system consisted of
two chiral scalar superfields $Z ^ \alpha $ ($ \alpha=1,2$)  satisfying
$$ \bar D_\pm Z ^ \alpha=0, \qquad D_\pm \bar Z ^{\bar \beta} =0
\eqn\chir$$
where $ +,-$ are chiral spinor indices. (The superspace conventions are as in
[\GHR].)
The lowest components of the superfield, $Z^\alpha \mid
 _{ \theta=0} = z^ \alpha$, are the bosonic complex coordinates of the
space-time.
The matter system is given by the N=2 sigma-model
$$
S= \int{d^2\sigma d^4 \theta \, K( Z, \bar Z)}
\eqn\zum$$
where $K$ is the Kahler potential, so that the metric is
$$ g_{\alpha \bar \beta}= {\partial^2 K\over \partial z^\alpha \partial  \bar z
^ {\bar \beta}}
\eqn\kah$$
For this to be a consistent string background, the sigma-model
 must be conformally invariant, which will be the case if the metric is
Ricci-flat, or equivalently
if the curvature is self-dual (or anti-self-dual)
$$*R_{\mu \nu \rho
\sigma}
 \equiv {1 \over 2}{ \epsilon _{\mu \nu} }^{\lambda \tau}
R_{\lambda\tau \rho \sigma} = \pm R_{\mu \nu \rho
\sigma}
\eqn\self$$
where $\mu,\nu\dots=1,\dots 4$ are coordinate  indices in a real coordinate
system.
As
$$R_{\alpha \bar \beta}= \partial_\alpha \partial _ \bb \log det g_{ \alpha
\bb}
\eqn\ris
$$
the Ricci-flatness condition can be integrated to give
$$det g_{ \alpha \bb}
= f(z)\bar f (\bar z)
\eqn\con$$
for some holomorphic function $f(z)$. With a suitable choice of coordinates,
the right hand side
can be set to $-1$   for (2,2) space-time signature or to $+1$ for (4,0)
signature.
 Writing $K=\tilde K + \phi$  where $\tilde K$ is a background potential,
this gives an equation for $\phi$
that can be derived from the Plebanski action (in the notation of [\Oo])
$$
\int
\partial \phi \bar \partial \phi + {1\over 3} \phi \partial \bar \partial  \phi
_   \wedge
\partial
\bar \partial
\phi
\eqn\pleb$$
In [\Oo], it was shown that the string spectrum consists of  a scalar field
$\phi $ whose
effective dynamics was found from
string scattering amplitudes  to be governed by precisely the   effective
action \pleb, confirming
that the N=2 string is a theory of self-dual gravity.

This is not the most general way in which  conformal  invariance can be
achieved. More generally,
one can introduce a dilaton $\Phi$ coupling through a supersymmetric
Fradkin-Tseytlin term so that
the condition for one-loop conformal invariance becomes
$$
R_{\mu \nu} = -2\nabla _\mu \nabla _\nu \Phi
\eqn\conr$$
which implies
$$\eqalign{&\nabla _\alpha \nabla _\beta \Phi=0
\cr &
\partial_\alpha \partial _ \bb \left( \log det g_{ \alpha \bb}-2\Phi \right)=0
\cr}
\eqn\modr$$
These imply that $\xi_\alpha =i\partial \Phi$ is a holomorphic Killing vector
with Killing
potential $\Phi$, which satisfies
$$\Phi = {1\over 2} \log det g_{ \alpha \bb}\eqn\cde$$
in a suitable coordinate system [\Kou].
In addition, there is a one-loop contribution to the Virasoro central charge
proportional to
$(\nabla \Phi)^2$, so that there are solutions in more than four dimensions.
This gives field equations for $\phi,\Phi$ which generalise the
self-duality equations described above; some solutions with non-trivial dilaton
are given in [\Kou].

However, the Kahler sigma-model is not the most general matter system with
(2,2) supersymmetry.
The aim of this paper is to consider $N=2$ strings based on the sigma-models of
[\GHR], using the
results of [\GHR-\HW]. The most general sigma-model with (2,2) supersymmetry
off-shell has both
chiral superfields
$U^a,\bar U^{\bar b} $ ($a,b =1,2,\dots , d_1$)
satisfying the constraints
$$
\bar D_\pm U^a=0, D_\pm \bar U^{\bar b} =0
\eqn\chiru$$
and twisted
chiral superfields $V^i, \bar V^{\bar j}$ ($i,j =1,2,\dots ,d_2$) satisfying
the constraints
$$D_+V^i=0,\qquad
\bar D_-V^i=0,\qquad
D_- \bar V^{\bar j} =0,
\qquad
\bar D_+\bar V^{\bar j}=0
\eqn\twi$$
The action
$$
S= \int{d^2\sigma d^4 \theta \, K( U, \bar U,V,\bar V)}
\eqn\twiact$$
defines a supersymmetric non-linear sigma-model with torsion on a  target space
of complex dimension
$d_1+d_2$ with coordinates $x^\mu=(u,\bar u,v,\bar v)$ ($\mu=1, ...
2(d_1+d_2)$) where  $u$ is
the lowest component of the superfield $U$ etc.
The bosonic part of the component sigma-model action
is $$
S= {1\over 2}\int d^2\sigma   \, (g_{\mu \nu }\partial _ a x^\mu \partial ^ a
x^\nu +
b_{\mu \nu }\epsilon ^{ab}\partial _ a x^\mu \partial _b x^\nu)
\eqn\comact$$
where the metric
$g_{\mu
\nu}$ and anti-symmetric tensor
$b_{\mu
\nu}$ whose curl defines the  torsion $H_{\mu \nu \rho}= {1\over 2} \partial
_{[ \mu }b_{\nu \rho
]}$ are given by
$$\eqalign{ g_{a \bar b}&=K_{a \bar b}, \qquad g_{i \bar j}=-K_{i \bar j}
\cr
b_{a \bar j}&=K_{a \bar j}, \qquad b_{i \bar b}=K_{i \bar b}
\cr}
\eqn\met$$
All other components of $g_{\mu \nu}$ and  $b_{\mu \nu}$ not related to these
by complex
conjugation or symmetry vanish, and $K_{\mu \nu ...\rho} $ denotes the partial
derivative
$\partial _\mu \partial _ \nu ...\partial _\rho K$. The geometry is that of a
hermitian locally
product space with two commuting complex structures ${J^{\pm \mu}}_\nu$; see
[\GHR] for details.
In the special cases in which  either $d_1=0$ or $d_2=0$, the torsion vanishes
and the space is
Kahler.
It is useful to define the connections with torsion
$$
{\Gamma ^\pm } _{\mu \nu }^\rho= {\Gamma  } _{\mu \nu }^\rho \mp H _{\mu \nu
}^\rho
\eqn\conn$$
where $\Gamma$ is the usual Christoffel connection. The complex structures are
each covariantly
constant with respect to the corresponding connection: $\nabla ^+ J^+=0, \nabla
^- J^-=0 $ [\GHR],
so that both connections have holonomy $	U(d_1+d_2)$.
It will also be useful to define the vectors $v^\pm$ by
$$ v^\pm _\mu =\pm J^ {\pm}_{\mu \nu }J^ {\pm}_{  \rho \sigma}H^{  \nu \rho
\sigma}
\eqn\viss$$
and the U(1) parts of the two curvature tensors $R^\pm_{\mu  \nu \rho \sigma}$
(defined as in [\Hsig])
$$C^\pm _{\mu \nu} =J^{\pm \rho \sigma} R^\pm_{\mu  \nu \rho \sigma}\eqn\cpm$$
It follows from \met\ that
$$\eqalign{
 v^\pm _a &=   \partial _ a \log det (g_{i \bar j})
\cr
v^\pm _i &=  \partial _ i \log det (g_{a \bar b})
\cr}
\eqn\vis$$

The sigma-model will be conformally invariant at one-loop if there is a scalar
$\Phi(x)$
such that
$$\eqalign{
R_{\mu \nu} -H_{  \mu}{}^{ \rho \sigma}H_{  \nu \rho \sigma}&= -2\nabla _\mu
\nabla _\nu \Phi
\cr
\nabla ^\nu H _{  \nu \rho \sigma}&= 2H_{  \nu \rho \sigma} \nabla ^\nu \Phi
\cr}
\eqn\reqmot$$
%%%%%
These are satisfied if [\Bus,\Hsig,\Roc]
$$
 \log det (g_{a \bar b} ) - \log det (g_{i \bar j}) = f(u) +\bar f(\bar u)
+g(v) +\bar g(\bar v)
\eqn\cmog$$
with $$ \Phi =- {1\over 2}\log det (g_{a \bar b} )\eqn\fis
$$
for some holomorphic functions $f,g$. With a suitable choice of coordinates,
the right hand side of
\cmog\ can be set to zero.  Then
$$v^\pm _\mu =  -2 \partial _\mu \Phi\eqn\vtert$$
and it follows from [\Hsig] that
$$C^\pm _{\mu \nu} =0\eqn\ciso$$
so that both connections $\Gamma ^\pm$ have $SU(d_1+d_2)$ holonomy and the
first Chern class
vanishes. Conversely, if \ciso\ holds, then using $C_{ia}= i\partial _i
v_a-i
\partial _a v_i$ and
\vis, it can be shown  that \cmog,\fis\ hold.

Thus if \ciso\ is satisfied, the sigma-model will be conformally invariant at
one loop, but
there will in general be conformal anomalies at four loops and higher [\GHL].
It is straightforward to show, using an argument similar to that of [\NS], that
a deformation of $K$
can be defined order by order in the sigma-model loop-counting parameter
($\hbar $ or $\alpha ^
\prime$) so that the beta-functions of the deformed sigma-model vanish.
Although for general
sigma-models this is not sufficient for conformal invariance [\HT], in this
case
it is possible to cancel the conformal anomaly to all orders by choosing the
dilaton to be given
again by
\fis, but now with the deformed metric appearing on  the right hand side.

The equations \reqmot\ imply that
$$
C \equiv -R + {1\over 3} H^2 -4 \nabla ^2 \Phi +4 (\nabla \Phi )^2
\eqn\cisss$$
is a constant.
The resulting conformal
field theory has central charge  given by
$$c=3(d_1+d_2)
+{1 \over 4} \alpha ^\prime C + O({\alpha ^\prime}^2)\eqn\cen$$
However, for (2,2) geometries, the identity
$$
\nabla ^\mu v^\pm _\mu =   (v^\pm )^2 - {2\over 3} H^2
\eqn\ids$$
implies that the quantity $C$ given by \cisss\
is given by
$$C=2 \nabla^\mu v^\pm_\mu= -\nabla ^2 \Phi=0\eqn\cmod$$
 when \reqmot\ holds.
In four dimensions, the definition \viss\ implies
$$ v^\pm _\mu =\pm {1\over 3} \epsilon  _{\mu \nu   \rho \sigma}H^{  \nu \rho
\sigma}
\eqn\visst$$
so that $\nabla^\mu v^\pm_\mu=0$ and $C$ vanishes identically. In this case,
the sigma-model has in
fact (4,4) supersymmetry [\GHR], the central charge is given exactly by
$c=3(d_1+d_2)$ and the theory is conformally invariant to all orders [\HP].
 In other dimensions, integrating \cmod\ and using the fact that $C$ is both
constant and a total
divergence  gives that $C=0$ for compact spaces, or for non-compact ones in
which
$\partial \Phi$ tends to zero sufficiently fast asymptotically.  In other cases
it can be non-zero,
so there are quantum corrections to the central charge.

These geometries are the generalisations of the Ricci-flat Kahler spaces. As in
the untwisted case,
the general solution of \reqmot\ involves Killing vectors.
Let
$$\eqalign{\xi_a&=i\partial _a\left[\log det (g_{i\bar j})-2\Phi\right]
\cr
\xi_i&=i\partial _i\left[\log det (g_{a\bar b})-2\Phi\right]
\cr}
\eqn\dfgjhd$$
Then \reqmot\ implies that $\xi_a$ and $\xi_i$ are two commuting Killing
vectors that are
holomorphic with respect to both complex structures and which satisfy
$$\partial _a \xi^i=0, \qquad \partial _i \xi^a=0
\eqn\comrt$$
A Killing vector $\xi^\mu$ will leave the torsion $H_{\mu \nu \rho}$ invariant
provided
$$H_{\mu \nu \rho}\xi^\rho=\partial _{[\mu} u_{\nu]}\eqn\reter$$
for some $u_\mu$. The equations \reqmot\
in addition imply that \dfgjhd\ satisfy \reter\ with
$$\eqalign{u_a&=i\partial _a\left[\log det (g_{a\bar b})-2\Phi\right]
\cr
u_i&=i\partial _i\left[\log det (g_{i\bar j})-2\Phi\right]
\cr}
\eqn\dfgjd$$
In general, $C$ will be non-vanishing and the central charge gets quantum
corrections, so that
solutions in dimensions other than four are possible. The previous case
\cmog,\fis\
arises when   the Killing vectors \dfgjhd\ both vanish.

Many geometries satisfying \cmog,\fis\ with vanishing  $\xi_i,\xi_a$  and
leading to conformally
invariant sigma-models are given implicitly by the twistor transform
construction of [\GHR] (these
in fact have (4,4) supersymmetry and so do not need quantum deformations of
$K$).
An explicit compact example with $\Phi=0$ and non-vanishing $\xi_i,\xi_a$
was given in [\RSS], while some non-compact examples both with and without
Killing vectors were
discussed in [\Kou].

To construct a (2,2) string theory, it is necessary to choose $c=6$ to cancel
the ghost
contributions to the anomalies.
The simplest case is that of   backgrounds with  $d_1+d_2=2$ and vanishing
$\xi_i,\xi_a$, so that
$C=0$. This gives   the Ricci-flat Kahler case
  if either $d_1$ or $d_2$ vanishes, while there will be non-vanishing
torsion only if $d_1=d_2=1$. Then the connections $\Gamma ^\pm$ both have SU(2)
holonomy so that the model
has (4,4) supersymmetry [\Van] and is perturbatively finite [\HP] and is in
fact conformally
invariant (as above).
Both curvatures with torsion are anti-self-dual on both the first two and last
two indices:
$${1 \over 2}{ \epsilon _{\mu \nu} }^{\lambda \tau}
R^{\pm }_{\lambda\tau \rho \sigma} ={1 \over 2} R^{\pm }_{\mu \nu \lambda\tau}
 {\epsilon^{\lambda\tau}}_{\rho\sigma}=-R^{\pm }_{\mu \nu \rho
\sigma}\eqn\curer$$
 as can be checked using \ciso, $J_{[\mu\nu}J_{\rho\sigma]}=-{1 \over
3}\epsilon_{\mu
\nu \rho \sigma} $ and [\Hsigb]
$$R^+_{\mu \nu \rho \sigma}=R^-_{ \rho \sigma\mu \nu}\eqn\reqa$$
In this case the equation \cmog\ reduces to the linear Laplace equation
$$ K_{u\bar u} + K_{v\bar v}=0
\eqn\lap$$
(choosing coordinates so that the right hand side vanishes)
which can be derived from the free effective action
$${1\over 2}\int d^4 x \, \partial _\mu K \partial ^\mu K
\eqn\free
$$
This should also be the effective action that arises from considering the
scattering
amplitudes for such (2,2) strings; it would be interesting to check this
explicitly.

To summarise, a (2,2) string theory can be constructed   using $d_1$
chiral multiplets and
$d_2$ twisted chiral multiplets, and the geometry is specified in terms of a
potential $K$ and a dilaton $\Phi$. In the general situation, the gradient of
the dilaton is given in terms of a Killing vector $\xi_\alpha$. The case in
which $\xi_\alpha=0$ is easiest to analyse; in this case
  $d_1+d_2=2$ and the   space-time is four-dimensional. If $d_1=2$ or $0$, then
the geometry is Kahler,
     $K$ is the Kahler potential and $\Phi=0$. The dynamics require the Ricci
tensor to vanish,
which is equivalent to requiring the curvature to be anti-self-dual. The matter
content is a scalar
field governed by the Plebanski action \pleb. If $d_1=d_2=1$, however, the
geometry is no longer
Kahler but is instead a complex geometry with torsion of the type introduced in
[\GHR] which is
again given
 in terms of a
potential $K$. The metric can have signature (2,2) or (4,0) and the dilaton is
given by \fis. The dynamics require the curvature with
torsion to satisfy the anti-self-duality conditions \curer. The string spectrum
should again be the
scalar field corresponding to the potential
$K$, but now the dynamics is given by the free action \free. The target space
theory should be a supersymmetric theory of self-dual supergravity in 2+2
dimensions; such theories have been studied in [\Gates,\Sez].
This exhausts the possibilities if the (2,2) world-sheet supersymmetry is to
close off-shell.
However, there remains the class of  on-shell (2,2) models with two complex
structures that do not
commute and it would be interesting to understand the resulting string theory
in this case also.
Similar results can also be derived for  for (2,1) strings and (2,0) strings.
In particular, these involve a four-dimensional space-time for which the
curvature with torsion has
$SU(2)$ holonomy, and so is self-dual [\Hsig-\Van]. This has particular
relevance for the proposal
of [\MK] relating the (2,1) string to the type IIB string and the $D=11$
membrane.  This will be
discussed in more detail elsewhere.

%%%%%%
%\vskip 0.5cm
\noindent{\bf Acknowledgements}: I would like to thank A. Sevrin for some
helpful comments.
%%%%%%%%%%%%%%%%%%%%%%%%%%%%%%%%%%%%%%%%%%%%%%%%
\refout
\bye